\documentclass[preprint,12pt]{elsarticle}
\journal{journal} \RequirePackage{graphicx}

\begin{document}

\begin{frontmatter}

\title{Anisotropic stars for spherically symmetric spacetimes satisfying the Karmarkar condition}

\author{S.K. Maurya}
\address{Department of Mathematical and Physical Sciences,
College of Arts and Science, University of Nizwa, Nizwa, Sultanate
of Oman\\sunil@unizwa.edu.om}

\author{B.S.Ratanpal}
\address{Department of Applied Mathematics, Faculty of Technology \& Engineering,The M. S. University of Baroda, Vadodara - 390 001, India\\bharatratanpal@gmail.com}

\author{M. Govender}
\address{Department of Mathematics, Faculty of Applied Sciences, Durban University of Technology, Durban, South Africa
\\megandhreng@dut.ac.za}

\date{Received: date / Accepted: date}

\maketitle

\hrule

{\bf H~I~G~H~L~I~G~H~T~S}

$*$ We consider the Einstein field equations for spherically symmetric spacetimes of embedding class one.

$*$ We explore the Karmarkar condition for spherically symmetric, bounded matter configurations.

$*$ The Karmarkar condition reduces the solution-generating method to the Einstein field equations to a single metric function.

$*$ By specifying this metric function we generate physically plausible stellar configurations with anisotropic pressure.

\hrule

\begin{abstract}
A new class of solution describing an anisotropic stellar configuration satisfying Karmarkar's condition i.e. spherically
symmetric metric of embedding class 1, is reported. It has been shown that the compact star model is physically well-behaved
and meet all the physical requirements for a stable configuration in hydrostatic equilibrium. Our model describes compact stars like Vela X-1 and 4U1608-52 to a very good approximation.
\end{abstract}


\end{frontmatter}

\section{INTRODUCTION:}

A century ago Karl Schwarzschild\cite{S1} obtained the first exact solution of the Einstein field equations. Since then many exact solutions of
the Einstein field equations have appeared in the literature, with only a small subclass representing physically viable stellar models. Schwarzschild's constant density sphere has been generalised to include physically observed phenomena such as the elctromagnetic field, pressure anisotropy, deviations from spherical symmetry, dissipation and rotation. A systematic and comprehensive study of solutions of the Einstein field equations was carried out by Delgaty and Lake\cite{LD}. They analysed 127 known exact solutions of Einstein's field equations out of which only nine solutions satisfy all the physical plausibility conditions. This shows the complexity in getting exact solutions of Einstein's
field equations describing physically realizable astrophysical objects. This has spurred researchers to search for solutions which are physically viable but more importantly, which are in good approximation to observational data.
\\\\
\noindent Ruderman\cite{R} and Canuto\cite{C} showed that when the matter density is much higher than the nuclear density,
matter may be anisotropic in nature. Bower and Liang\cite{BL} and Herrera and Santos\cite{HS} carried out extensive studies on the impact of anisotropy on self-gravitating configurations. The anisotropy may occur due to presence of type 3A superfluids\cite{R}\cite{BL}\cite{KW},
phase transitions\cite{SMK} within the core or due to electromagnetic fields\cite{KMSM}.
\\\\
\noindent Vaidya and Tikekar\cite{VT}, Tikekar and Thomas\cite{TT1} and Tikekar \& Jotania\cite{TJ} studied models of relativistic
stars on spheroidal, pseudo-spheroidal and paraboloidal spacetimes respectively. Charged stars on spheroidal spacetime have been
studied by Patel and Kopper\cite{PK}, Sharma {\em et. al.}\cite{SMM}, Gupta and Kumar\cite{GK} \& Komatiraj and Maharaj\cite{KM}.
The compact objects on pseudo-spheroidal spacetime have been studied by Tikekar and Thomas\cite{TT2}, Thomas {\em et. al.}\cite{TRV}
\& Chattopadhyay and Paul\cite{CP}. The core envelope models on pseudo-spheroidal spacetimes have been studied by Thomas and Ratanpal\cite{TR}.
The paraboloidal spacetime is a particular case of the Finch and Skea\cite{FS} spacetime. The relativistic star model admitting quadratic
equation of state on paraboloidal spacetime was studied by Sharma and Ratanpal\cite{SR}. These studies suggest that geometrically
significant spacetimes can be used to describe the physically realistic stars.
\\\\
\noindent The embedding problem is one of the interesting problems on geometrically significant spacetimes which was first addressed by  Schlai\cite{S}. Nash\cite{N} provided the first isometric embedding theorem. Karmarkar\cite{K} derived the condition for
embedding 4-dimensional spacetime metric in 5-dimensional Euclidean space. Karmarkar classified these spacetimes as class-1 spacetime.
For a spherically symmetric spacetime metric, the Karmarkar condition in terms of curvature components takes the form
\begin{equation}\label{Kcondition}
R_{1414}R_{2323}=R_{1212}R_{3434}+R_{1224}R_{1334}.
\end{equation}

\noindent Recently Karmarkar's condition attracted attention amongst many researchers working on exact solutions of the Einstein field equations, modeling compact objects and stability analyses of self-gravitating objects. Maurya {\em et. a.}\cite{MGRC} began with the study of charged
compact stars satisfying Karmarkar's condition. This led to a flourishing of models of compact objects satisfying the Karmarkar condition\cite{MGDR}-\cite{SP}. In the present work we have considered the
spherically symmetric spacetime metric of embedding class 1 and obtained the singularity-free solution of Einstein's field equations for an
anisotropic fluid distribution. We have shown that the model satisfies all the physical plausibility conditions and is also stable.
The work is organized as follows: Section 2 introduces the Einstein's field equations, TOV equation and Karmarkar's condition for spherically
symmetric spacetimes necessary for this investigation. We present the anisotropic solution of embedding class one for compact stars in section 3. We consider the matching conditions of the interior spacetime to the vacuum Schwarzschild exterior solution in section 4.  The physical features of our model are discussed in section 5. We conclude with a discussion of our results in section 6.

\section{Einstein field equations, TOV equation and Karmarkar condition for spherical symmetric metric:}
\subsection{Einstein field equations}

\noindent We begin with the static spherically symmetric spacetime metric given by
\begin{equation}
ds^{2} = -r^{2} (d\theta ^{2} +\sin ^{2} \theta \, d\phi ^{2} )-e^{\lambda(r) } dr^{2}+e^{\nu(r) } \, dt^{2},
\label{1}
\end{equation}
\noindent where, $e^{\lambda(r)}$ and $e^{\nu(r)}$ represent the gravitational potential for the interior anisotropic fluid distribution.
The energy-momentum tensor for the anisotropic fluid distribution has the form
\begin{equation}
\label{3}
T^{i}_{j}=(\rho+p_t)\,v^i\,v_j-p_t\,g^i_j+(p_r-p_t)\,u^i\,u_j,
\end{equation}
\noindent where $p_r$, $p_t$ and $\rho$ denote the radial pressure, tangential pressure and matter density respectively. The contravariant components $v^i$ is the velocity four vector and $u^i$ is the unit space-like vector in the radial direction.
With the metric (\ref{1}) together with the energy-momentum (\ref{3}) Einstein's field equations take the form
\begin{equation}
\label{4}
p_{r} =\frac{e^{-\lambda}}{8\pi}\left[\frac{v'}{r}  -\frac{(e^{\lambda}-1)}{r^{2}} \right]  ,
\end{equation}
\begin{equation}
\label{5}
p_{t} =\frac{e^{-\lambda}}{8\pi}\left[\frac{v''}{2} -\frac{\lambda'v'}{4} +\frac{v'^{2} }{4} +\frac{v'-\lambda'}{2r} \right] ,
\end{equation}
\begin{equation}
\label{6}
\rho =\frac{e^{-\lambda}}{8\pi}\left[\frac{\lambda '}{r} +\frac{(e^{\lambda}-1 )}{r^{2}}\right].
\end{equation}

\noindent Here primes denote the derivative with respect to the radial coordinate $r$. The value of gravitational constant and velocity of
light are taken to be unity in above coupled differential equations. Using Eqs.(4) and (5) we obtain the anisotropic factor

\begin{equation}
\label{7}
\Delta =\,p_{t} -\, p_{r} \,=\frac{e^{-\lambda }}{8\pi}\left[\frac{v''}{2} -\frac{\lambda 'v'}{4} +\frac{v'^{2} }{4} -\frac{v'+\lambda '}{2r} +\frac{e^{\lambda }-1}{r^{2} } \right],
\end{equation}
which vanishes when the pressure is isotropic and is zero at the centre of the fluid distribution.

\subsection{Tolman-Oppenhiemer-Volkoff (TOV) equation:}

From Eqs.(\ref{4}) and (\ref{6}) we can write

\begin{equation}
8\,\pi\,(\rho+p_r)= \frac{\lambda'+\nu'}{r}\,e^{-\lambda} \label{rop1},
\end{equation}

\begin{equation}	 8\,\pi\,\frac{dp_r}{dr}=\left[\frac{\nu''}{r}-\frac{\nu'\,\lambda'}{r}-\frac{\nu'}{r^2}-\frac{\lambda'}{r^2}\right]\,e^{-\lambda}+\frac{2(1-e^{-\lambda})}{r^3}.
\label{dpr1}
\end{equation}

\noindent Using Eqs.(\ref{4},\ref{5},\ref{rop1}) and (\ref{dpr1}) we get

\begin{equation}
\frac{2}{r}(p_t-p_r)=\frac{dp_r}{dr}+\frac{1}{2}\,\nu'\,(\rho+p_r)=0. \label{TOV1}
\end{equation}

\noindent The gravitational mass within a sphere of radius $r$ is derived from the Tolman-Whittaker formula
\begin{equation}
M_G(r)=\frac{1}{2}r^2e^{\frac{\nu-\lambda}{2}}\nu'. \label{Mg1}
\end{equation}

\noindent By plugging the value of $\nu'$ from Eq.(\ref{Mg1}) into Eq.(\ref{TOV1}) we get,

\begin{equation}
\frac{2}{r}(p_t-p_r)  -  \frac{dp_r}{dr}+\frac{M_G(r)\,(\rho+p_r)}{r^2}\,e^{\lambda-\nu}=0.
\end{equation}

\noindent The above equation represents the well-known generalized Tolman-Oppenheimer-Volkoff (TOV) equation which provides the equilibrium condition for an anisotropic stellar system.

\subsection{Karmarkar condition:}

In general, the spherically symmetric spacetime metric (\ref{1}) is of class two.
If the metric (\ref{1}) satisfies the Karmarkar condition (\ref{Kcondition}) it will then represent a spacetime of embedding class one. The components of the Riemann curvature tensor $R_{hijk}$ for metric (1) are given as:\\

$R_{2323}=\frac{sin^2\theta\,(e^{\lambda}-1)\,r^2}{e^{\lambda}}$, \,\, $R_{1212}=\frac{\lambda'\,r}{2}$,\,\,  $R_{2424}=\frac{\nu'\,r\,e^{\nu-\lambda}}{2}$,\\

$R_{1224}=0$,\,\,$R_{1414}=\frac{e^\nu}{4}\,[2\,\nu''+\nu'^2-\lambda'\,\nu']$,\,\, \,$R_{3434}=sin^2{\theta}\,R_{2424}$.\\

\noindent By inserting the components of $R_{hijk}$ into the Karmarkar condition (\ref{Kcondition}) we obtain the following differential equation,

\begin{equation}
\label{9}
\frac{\nu''}{\nu'}+\frac{\nu'}{2}=\frac{\lambda'\,e^{\lambda}}{2\,(e^{\lambda}-1)}.
\end{equation}

\noindent which is readily solved to give the gravitational potential $\nu$,

\begin{equation}
\label{10}
\nu=2\,ln\left[A_1+B_1\int{\sqrt{(e^{\lambda(r)}-1)}dr}\right].
\end{equation}
where,  $A_1$ and $B_1$  are non-zero arbitrary constant of integration.

By inserting Eq.(\ref{10}) into the Eq.(\ref{7}) and rearranging the terms we can recast $\Delta$ as

\begin{equation}
\label{D11}
\Delta=\frac{\nu'\,e^{-\lambda}}{32\,\pi}\,\left(\frac{\nu'\,e^{\nu}}{2B^{2}r}-1\right)\,\left(\frac{2}{r}-\frac{\lambda'e^{-\lambda}}{1-e^{-\lambda}}\right).
\end{equation}

\noindent The pressure anisotropy $\Delta$ is zero throughout the distribution if either first factor or second factor or both the factors on the right
side of (\ref{D11}) are zero. When the first factor on the right side of (\ref{D11}) is zero we get the Kohler-Chao\cite{KL}
solution while the vanishing of the second factor on the right side of (\ref{D11}) admits the Schwarzschild's\cite{S2} interior solution. The Kohler-Chao solution is cosmological nature as there is no finite radius for which the radial pressure vanishes. The interior Schwarzschild solution has several short-comings in modeling a stellar object, the most notable being infinite sound speeds within the core.

\section{Anisotropic solution of embedding class one for compact star:}
It is interesting to note that the solution of Einstein field equations for anisotropic matter distribution depends upon one of the metric functions $\nu$ or $\lambda$ because the Karmarkar condition gives a direct relation between the metric functions (for more details see the following references \cite{MGRC,MGDR}). For this purpose we make the following ansatz for $e^{\lambda}$,
\begin{equation}
e^{\lambda}=\frac{4 + cr^2\,\left[e^{(ar^2+b)}-e^{-(ar^2+b)}\right]^2}{4},  \label{lambda1}
\end{equation}

\noindent where, $a\ne0$, $b\ne0$ or $c\ne0$. If $a=b=0$ or $c=0$ then the spacetime takes the following form
\begin{equation}\label{flate}
ds^{2} =-dr^{2} -r^{2} (d\theta ^{2} +\sin ^{2} \theta \, d\phi ^{2} )+e^{\nu(r) } \, dt^{2},
\end{equation}

\noindent for which Karmarkar's condition (\ref{Kcondition}) is satisfied but spacetime metric (\ref{flate}) is not of class 1 as shown by
Pandey and Sharma\cite{PS}, hence we take $a$, $b$ and $c$ as positive constants. The metric potential $e^{\lambda}$ chosen here
does not give rise to spheroidal, pseudo-spheroidal or paraboloidal spacetimes. Also it does not represent the spheroidal
geometry considered by Schwarzschild or Kohler-Chao.

\noindent Substituting Eq.(\ref{lambda1}) into Eq.(\ref{10}), we get

\begin{equation}
\nu=2\,ln\left[A+B\left(\frac{e^{(ar^2+b)}+e^{-(ar^2+b)}}{2}\right)\right],  \label{nu1}
\end{equation}

\noindent where $A=A_1$ and $B=\frac{\sqrt{c}\,B_1}{2\,a}$ are constants .

%%%%%%%%%%%%%%%%%%%%%%%%%%%%%%%%%%%%%%%%%%%%%%%%%%%%%%%%%%%%%%%%%%%%%%%%%%%%%%%%%%%%%%%%%%%%%%%%%%%%
\begin{figure}[h!] \centering
	\includegraphics[width=5cm]{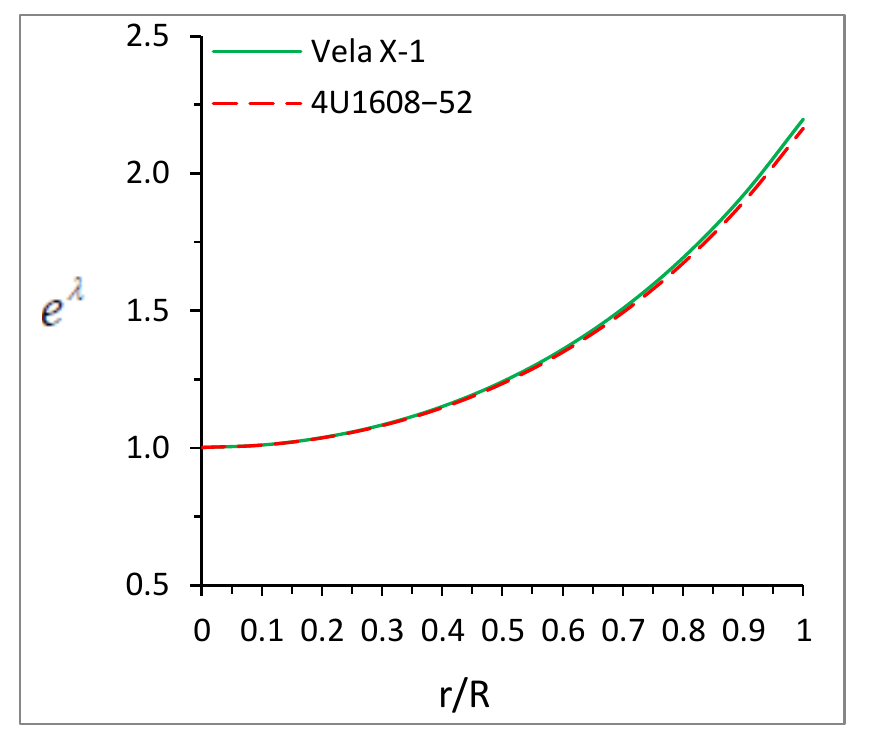} \includegraphics[width=5cm]{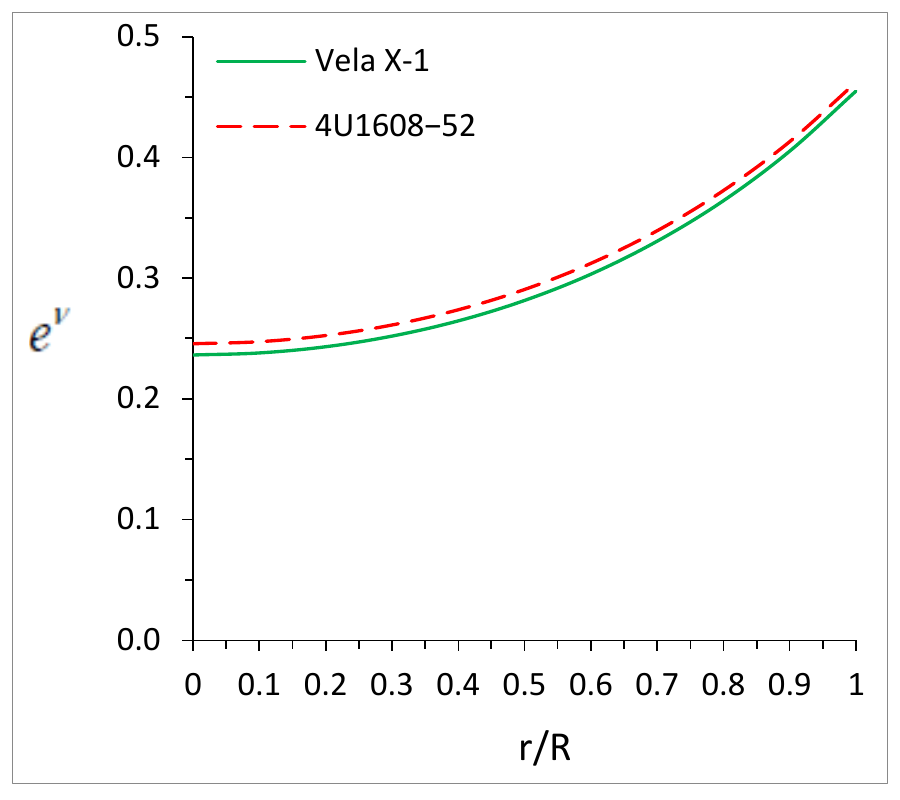}
	\caption{Behavior of the metric function $e^{\lambda}$ (left panel) and $e^{\nu}$ (right panel) against $r/R$ for the stars Vela X-1 and 4U1608-52. The numerical values for plotting this graph are as follows: (i). $a=0.00033$, $b = 0.19$, $c = 0.2663$, $R= 9.56 km$ and $M=1.77 M_{\odot}$ for Vela X-1 (ii). $a = 0.000325$, $b = 0.1845$, $c = 0.276$, $R=9.528$ and $M=1.74 M_{\odot}$ for 4U1608-52.}
	\label{11}
\end{figure}
%%%%%%%%%%%%%%%%%%%%%%%%%%%%%%%%%%%%%%%%%%%%%%%%%%%%%%%%%%%%%%%%%%%%%%%%%%%%%%%%%

\noindent For a physically viable model the metric functions $e^{\lambda}$ and $e^{\nu}$ must be finite at the centre while both should be
monotonically increasing functions of $r$. We observe from eq.(\ref{lambda1},\ref{nu1}), $(e^{\lambda})_{r=0}=1$ and $(e^{\nu})_{r=0}=\left[A+B\,\frac{e^b+e^{-b}}{2}\right]^2$, which are finite and free from singularity. Also Fig. (1) shows that $e^{\lambda}$ and $e^{\nu}$ both are increasing with $r$.\\

\noindent From the Eqs.(\ref{4},\ref{5},\ref{6}) together with Eqs.(\ref{lambda1},\ref{nu1}), we obtain $p_r$, $p_t$, $\rho$ and $\Delta$, (by taking\,
$x=a\,r^2+b$,\,\,$sinh(nx)=\frac{e^{nx}-e^{-nx}}{2}$,\,and\,  $cosh(nx)=\frac{e^{nx}+e^{-nx}}{2}$), as:

\begin{equation}
8\pi\,p_r=-\frac{sinhx[-8\,a\,B+2\,c\,A\,sinhx+c\,B\,sinh2x]}{2\,[A+B\,coshx]\,[1+c\,r^2\,sinh^{2}x]},
\end{equation}

\begin{equation}
8\pi\,p_t=\frac{2\,c\,A+(c+16a^2r^2)B\,coshx-2c\,A\,cosh2x-c\,B\,cosh3x+\Psi(x)}{4\,[A+B\,coshx]\,[1+c\,r^2\,sinh^{2}x]^2},
\end{equation}

%%%%%%%%%%%%%%%%%%%%%%%%%%%%%%%%%%%%%%%%%%%%%%%%%%%%%%%%%%%%%%%%%%%%%%%%%%%%%%%%%%%%%%%%%%%%%%%%%%%%
\begin{figure}[h!] \centering
	\includegraphics[width=5cm]{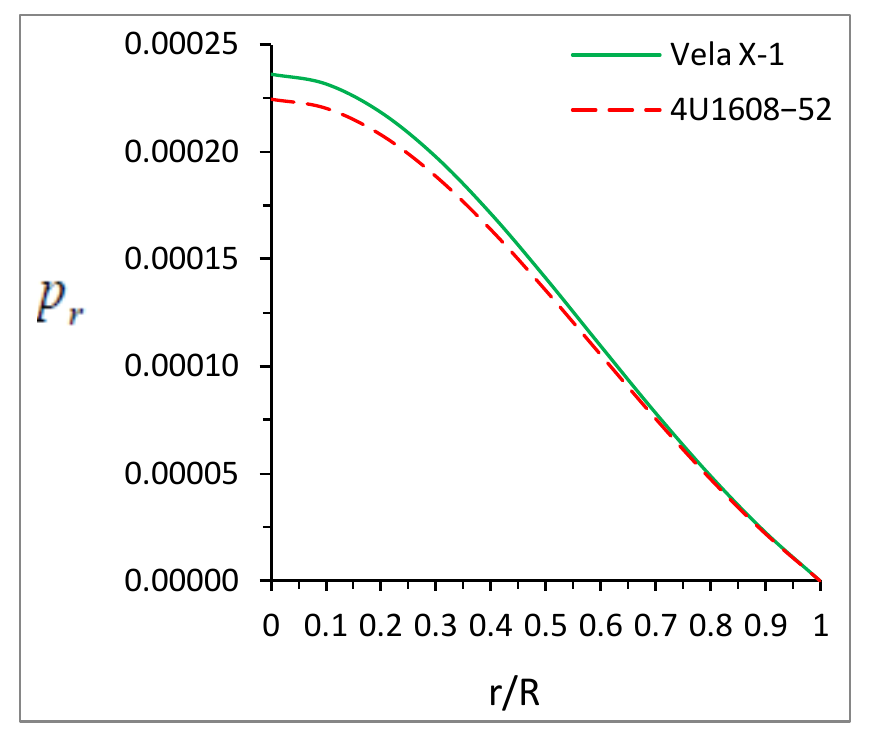} \includegraphics[width=5cm]{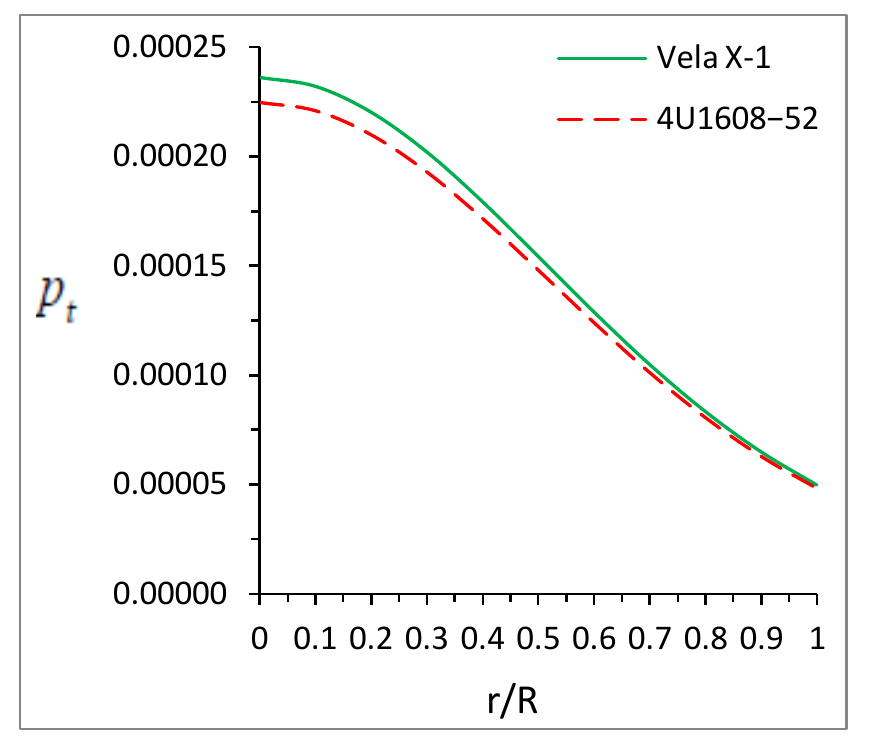}
	\caption{variation of radial pressure $p_r$ (left panel) and tangential pressure $p_t$ (right panel) against $r/R$ for the stars Vela X-1 and 4U1608-52. For this graph we have employed numerical values for $a$, $b$, $c$, $A$ and $B$ same as used in Fig.1  (see table 1).}
	\label{11}
\end{figure}
%%%%%%%%%%%%%%%%%%%%%%%%%%%%%%%%%%%%%%%%%%%%%%%%%%%%%%%%%%%%%%%%%%%%%%%%%%%%%%%%%

\begin{equation}
8\pi\,\rho=\frac{c\,[3\,sinh^2x+c\,r^2\,sinh^4x+2\,a\,r^2\,sinh2x]}{[1+c\,r^2\,sinh^{2}x]^2},
\end{equation}

%%%%%%%%%%%%%%%%%%%%%%%%%%%%%%%%%%%%%%%%%%%%%%%%%%%%%%%%%%%%%%%%%%%%%%%%%%%%%%%%%%%%%%%%%%%%%%%%%%%%
\begin{figure}[h!] \centering
	\includegraphics[width=5cm]{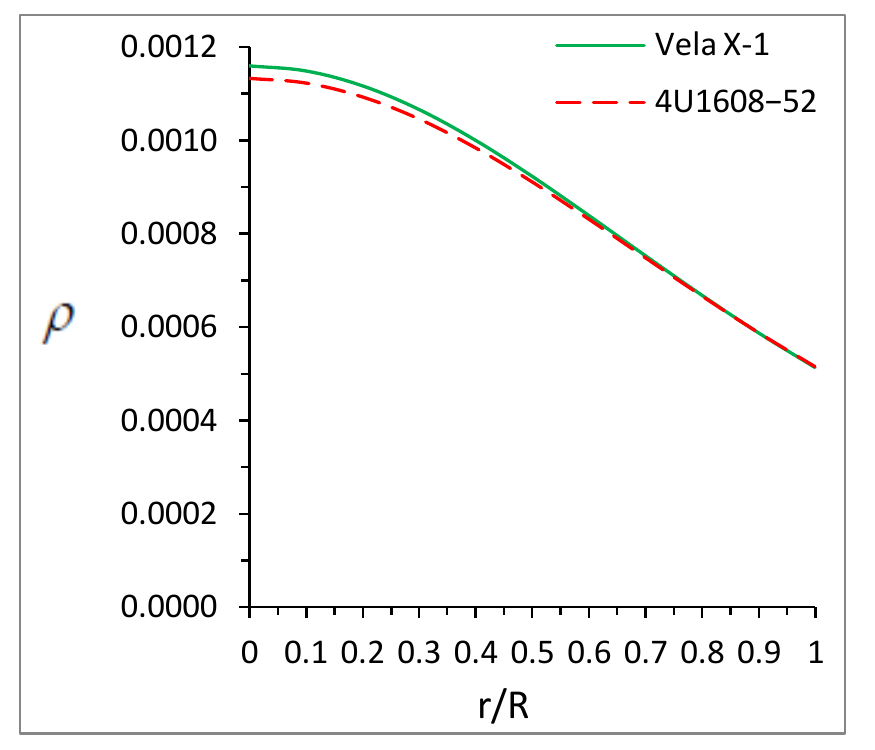}
	\caption{variation of energy density $\rho$ against $r/R$ for the star Vela X-1 and 4U1608-52. For this graph we have employed numerical values for $a$, $b$, $c$, $A$ and $B$ same as used in Fig.1 and 2  (see table 1).}
	\label{11}
\end{figure}
%%%%%%%%%%%%%%%%%%%%%%%%%%%%%%%%%%%%%%%%%%%%%%%%%%%%%%%%%%%%%%%%%%%%%%%%%%%%%%%%%

\begin{equation}
8\pi\,\Delta=\frac{r^2\,(c\,sinh^3x-2\,a\,coshx)\,(-4\,a\,B+2\,A\,c\,sinhx+c\,B\,sinh2x)}{2\,[A+B\,coshx]\,[1+c\,r^2\,sinh^{2}x]^2},
\end{equation}

\noindent where,\,\, $\Psi(x)=8\,a\,(2-c\,r^2)B\,sinhx-4ac\,A\,r^2\,sinh2x$. \\
The pressure anisotropy $\Delta$ is zero at centre $r=0$. However it can be made zero everywhere inside the star only when $c=0$ (which implies $B=0$). In this situation the metric turns out to be flat and all the physical parameters such as the radial pressure, tangential pressure and density vanish. Fig.(4) indicates that the anisotropy parameter is positive at each interior point of the matter configuration, ie., $p_t > p_r$. This indicates that the force due to local anisotropy is repulsive and may lead to more massive, stable configurations.

%%%%%%%%%%%%%%%%%%%%%%%%%%%%%%%%%%%%%%%%%%%%%%%%%%%%%%%%%%%%%%%%%%%%%%%%%%%%%%%%%%%%%%%%%%%%%%%%%%%%
\begin{figure}[h!] \centering
	\includegraphics[width=5cm]{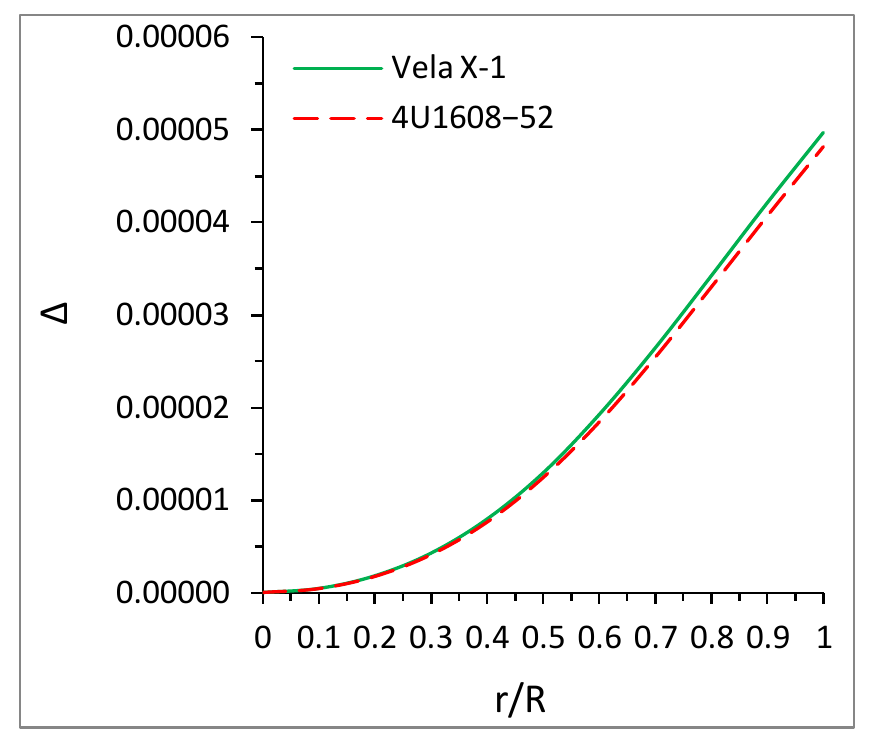}
	\caption{variation of anisotropic factor $\Delta$ versus $r/R$ for the star Vela X-1 and 4U1608-52. The numerical values for plotting this graph are as follows: (i). $a=0.00033$, $b = 0.19$, $c = 0.2663$, $R= 9.56 km$ and $M=1.77 M_{\odot}$ for Vela X-1 (ii). $a = 0.000325$, $b = 0.1845$, $c = 0.276$, $R=9.528$ and $M=1.74 M_{\odot}$ for 4U1608-52.}
	\label{11}
\end{figure}
%%%%%%%%%%%%%%%%%%%%%%%%%%%%%%%%%%%%%%%%%%%%%%%%%%%%%%%%%%%%%%%%%%%%%%%%%%%%%%%%%

\subsection{Bound on the constants:}

\subsubsection{Regularity of $p_r$, $p_t$ and $\rho$ at centre: }

The central pressures and central density are given as:

\begin{equation}
p_{r0}=p_{t0}=\frac{(e^b-e^{-b})\,[16\,a\,B-2\,c\,A\,(e^b-e^{-b})-c\,B\,(e^{2b}-e^{-2b})]}{32\,\pi\,[A+B\,(e^b+e^{-b})]},
\end{equation}

\begin{equation}
\rho_0=\frac{3\,c\,(e^b-e^{-b})^2}{32\,\pi},
\end{equation}

\noindent where, $sinh(nb)=\frac{e^{nb}-e^{-nb}}{2}$, \,\, $cosh(nb)=\frac{e^{nb}+e^{-nb}}{2}$, since central pressures are positive we obtain

\begin{equation}
\frac{A}{B} < \frac{16a-c\,(e^{2b}-e^{-2b})}{2\,c\,(e^b-e^{-b})}. \label{AB1}
\end{equation}

\noindent Here central density is positive as $a$, $b$ and $c$ already considered to be positive.

\subsubsection{Zeldovich's condition:}

Zeldovich's condition $p_{r}/\rho_0$ and $p_{t}/\rho_0$ must be $\leq 1$ at centre, places the following restriction on the constants

\begin{equation}
\frac{16a-4c\,(e^{2b}-e^{-2b})}{5\,c\,(e^b-e^{-b})} \leq \frac{A}{B} \label{AB2}.
\end{equation}

\noindent By using Eqs. (\ref{AB1}) and (\ref{AB2}) we get the following inequality:

\begin{equation}
\frac{16a-4c\,(e^{2b}-e^{-2b})}{5\,c\,(e^b-e^{-b})} \leq \frac{A}{B}<\frac{16a-c\,(e^{2b}-e^{-2b})}{2\,c\,(e^b-e^{-b})}.
\end{equation}

\section{Boundary conditions for the solution:}

The obtained interior solution must match continuously with the Schwarzschild exterior solution

\begin{equation}
ds^{2} = -r^{2} (d\theta ^{2} +\sin ^{2} \theta \, d\phi ^{2} )-\left(1-\frac{2M}{r} \right)^{-1} dr^{2}+\left(1-\frac{2M}{r} \right)\, dt^{2} ,
\end{equation}

\noindent at the boundary of stellar configuration $r=R$, where  $M$ is total mass of anisotropic stellar configuration contained within a sphere of radius $R$.
\noindent By matching the first fundamental form (continuity of $e^{\nu}$ and $e^{\lambda}$) and second fundamental forms (continuity of $\frac{\partial g_{tt}}{\partial r}$ i.e. $(p_r)_R=0$ ) of the interior solution with exterior Schwarzschild solution at the boundary of the star ($r=R$), we get (by taking $X=aR^2+b$):

\begin{equation}
1-\frac{2M}{R} =e^{\nu_R}=\left[A+B\left(\frac{e^{X}+e^{-X}}{2}\right)\right]^2,
\end{equation}

\begin{equation}
1-\frac{2M}{R} =e^{-\lambda_R}=\frac{4}{2(2-cR^2)+cR^2\,[e^{2X}-e^{-2X}]},
\end{equation}

\begin{equation}
(p_r)_R=0.
\end{equation}

\noindent By solving the above boundary conditions we obtain the constants as:

\begin{equation}
A=\frac{2\,[16a-c\,(e^{2X}-e^{-2X})]}{8\,a\,\sqrt{4+cR^2\,(e^X-e^{-X})^2}},
\end{equation}

\begin{equation}
B=\frac{\,c\,(e^{X}-e^{-X})}{4\,a\,\sqrt{4+cR^2\,(e^X-e^{-X})^2}\,},
\end{equation}

\begin{equation}
M=\frac{R}{2}\,\left[\frac{cR^2\,[e^X-e^{-X}]^2}{4+cR^2\,[e^X-e^{-X}]^2}\right].
\end{equation}

\section{Physical features of the model}

\subsection{Equilibrium condition:}

For a system which has just left hydrostatic equilibrium on a time-scale comparable to the relaxation time of the fluid, the 'force side' of the Tolman-Oppenheimer-Volkoff (TOV) equation can be written as
\begin{equation}\label{tov1}
R = \frac{dp_r}{dr} + \frac{\nu'}{2}(\rho+p_r) - \frac{2}{r}(p_t-p_r),
\end{equation}
where $R$ represents the total force acting on a fluid element. If $R </>0$, then the force is directed $outwards/outwards$ within the fluid sphere. In the case of $R = 0$, the fluid sphere is said to be in hydrostatic equilibrium. It has been shown that the effective inertial mass density ($\rho + p_r$) is sensitive to the temperature and thermal conductivity of a fluid sphere in quasi-static equilibrium. This implies that sources of anisotropy (shear viscosity and density inhomogeneities) as well as dissipation in the form of heat flow play important roles in determining the final static configuration.
\noindent The TOV equation for a body in hydrostatic equilibrium can be written in equivalent form describing three different forces acting within the fluid distribution, viz., anisotropic force $F_a$, hydrostatic force $F_h$ and gravitational force $F_g$ such that,
\begin{equation}
F_a+F_h+F_g=0,
\end{equation}

\noindent where $F_a,F_h$ and $F_g$ are given as,
\begin{eqnarray}
F_a &=& \frac{2}{r}(p_t-p_r),\\
F_h &=& -\frac{dp_r}{dr},\\
F_g &=& -\frac{\nu'}{2}(\rho+p_r)
\end{eqnarray}
For our model we obtain
\begin{equation}
F_a=\frac{r\,(2\,a\,coshx-c\,sinh^3x)\,(4\,a\,B-2\,A\,c\,sinhx-c\,B\,sin2x)}{8\,\pi\,[A+B\,coshx]\,[1+c\,r^2\,sinh^{2}x]^2},
\end{equation}

\begin{eqnarray}
F_g= -\frac{2\,r\,a\,B\,sinh^2x\,[4\,a\,B+4\,a\,c\,A\,r^2\,coshx+F_{g1}(x)]}{8\,\pi\,[A+B\,coshx]^2\,[1+c\,r^2\,sinh^{2}x]^2},
\end{eqnarray}

\noindent where,
$F_{g1}(x)=4\,a\,B\,c\,r^2\,cosh2x+2\,c\,A\,sinhx+c\,B\,sinh2x$.

%%%%%%%%%%%%%%%%%%%%%%%%%%%%%%%%%%%%%%%%%%%%%%%%%%%%%%%%%%%%%%%%%%%%%%%%%%%%%%%%%%%%%%%%%%%%%%%%%%%%
\begin{figure}[h!] \centering
	\includegraphics[width=5.5cm]{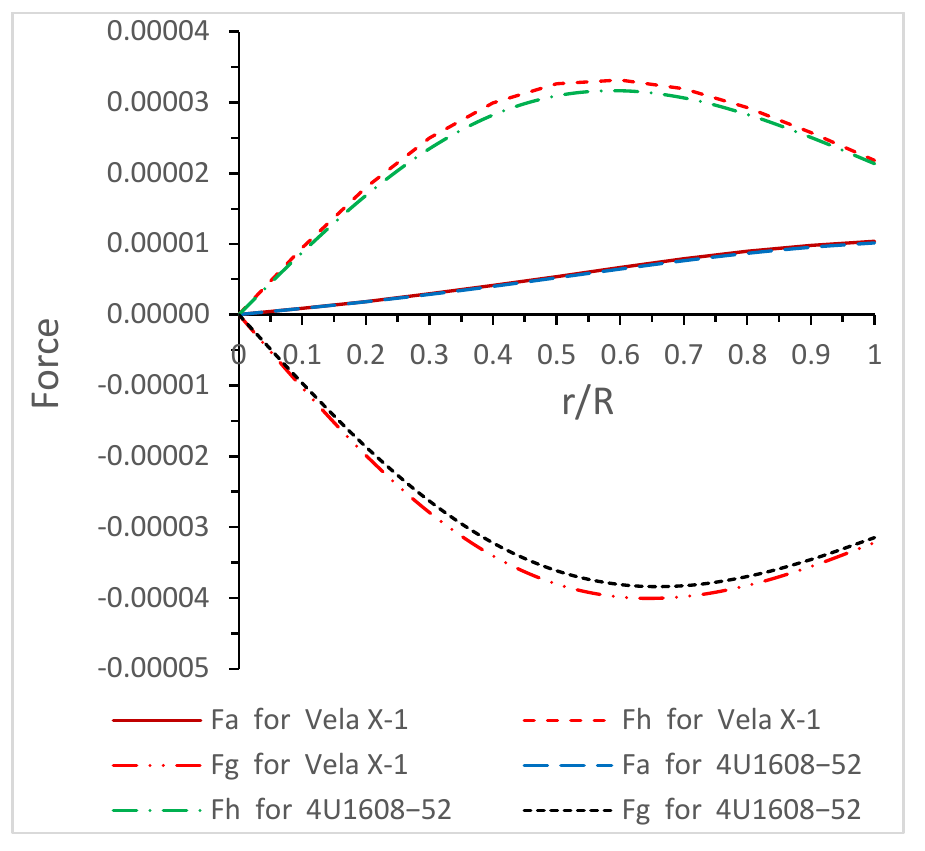}
	\caption{variation of different forces versus $r/R$ for the anisotropic star Vela X-1 and 4U1608-52. The numerical values for plotting this graph are as follows: (i). $a=0.00033$, $b = 0.19$, $c = 0.2663$, $R= 9.56 km$ and $M=1.77 M_{\odot}$ for Vela X-1 (ii). $a = 0.000325$, $b = 0.1845$, $c = 0.276$, $R=9.528$ and $M=1.74 M_{\odot}$ for 4U1608-52.}.
	\label{11}
\end{figure}

From  Fig.(\ref{11}), we note that the forces sum to zero at each interior point of the respective configurations. We also note that there are differences in magnitudes of the relative forces for each object. For example, the force due to anisotropy is smaller in 4U1608-52 (smaller mass) compared to Vela X-1 (larger mass). This is true for $F_g$ and $F_h$. Phenomenologically, this means that each fluid element in the more massive object is subject to higher hydrostatic force, gravitational force and the force due to anisotropy compared to its less massive counterpart, with the net sum of these forces being zero in each of these bodies.
%%%%%%%%%%%%%%%%%%%%%%%%%%%%%%%%%%%%%%%%%%%%%%%%%%%%%%%%%%%%%%%%%%%%%%%%%%%%%%%%%

\subsection{Stability criterion via cracking}

Herrera proposed the concept of cracking in a fluid sphere which has just lost hydrostatic equilibrium\cite{herher}. Any changes in the fluid distribution (density and pressure perturbations) are on a time scale comparable to the relaxation time. It is hypothesized that nonvanishing radial forces closer to the core directed inwards, may change sign at some point of the fluid sphere. Abreu et al.\cite{Ab} have shown that unstable regions may develop within the body when the tangential sound speed ($\partial{p_r}/\partial{\rho}$) exceeds the radial sound ($\partial{p_t}/\partial{\rho}$) speed. In order to achieve stable regions we must have
\begin{equation} \label{square}
-1 <  v_{t}^2 - v_{r}^2 \leq 0.\end{equation}
 For our solution we determine the square of radial and tangential velocity of sound as,

\begin{eqnarray}
v^2_r=\frac{dp_r}{d\rho}=\frac{dp_r/dr}{d\rho/dr}, \,\,\,\, v^2_t=\frac{dp_t}{d\rho}=\frac{dp_t/dr}{d\rho/dr},
\end{eqnarray}

\noindent where,

\begin{equation}
\frac{dp_r}{dr}=\frac{2\,r\,[\Psi_{r1}(x)+\Psi_{r2}(x)+(A+B\,coshx)\,\Psi_{r3}(x)+\Psi_{r4}(x)]}{16\,\pi\,\,(A+B\,coshx)^2 \,[1+c\,r^2\,sinh^2x]^2},
\end{equation}

\begin{equation}
\frac{dp_t}{dr}=-\frac{2\,r\,[\Psi_{t1}(x)\,\Psi_{t4}(x)+\Psi_{t2}(x)\,\Psi_{t6}(x)+\Psi_{t3}(x)\,\Psi_{t6}(x)]}{32\,\pi\,(A+B\,coshx)^2 \,[1+c\,r^2\,sinh^2x]^3},
\end{equation}

\begin{equation}
\frac{d\rho}{dr}=-\frac{2\,r\,c\,[6\,a\,c\,r^2\,coshx\,sinh^3x+5\,c\,sinh^4x+c^2\,r^2\,sinh^6x+\Psi_{d1}(x)]}{8\,\pi\,[1+c\,r^2\,sinh^2x]^3},
\end{equation}

with,

$\Psi_{r1}(x)=-2\,a\,c\,(A+B\,coshx)\,(A\,coshx +B\,cosh2x)\,sinhx\,[1+c\,r^2\,sinh^2x]$,

$\Psi_{r2}(x)=a\,coshx\,(A + B\,coshx)\, [1+ c\,r^2\,sinh^2x] [8\,a\,B-2\,c\,A\,sinhx-c\,B\,sinh2x]$,

$\Psi_{r3}(x)=c\,sinh^2x\,(2\,a\,r^2\,coshx+sinhx)\,[-8\,a\,B+2\,A\,c\,sinhx+c\,B\,sinh2x]$,

$\Psi_{r4}(x)=a\,B\,sinh^2x\,[1+c\,r^2\,sinh^2x]\,[-8\,a\,B+2\,c\,A\,sinhx+c\,B\,sinh2x]$,

$\Psi_{t1}(x)=a\,(A + B\,coshx) [1+ c\,r^2\,sinh^2x]$,

$\Psi_{t2}(x)=2\,c\,(A+B\,coshx)\,sinhx\,(2\,a\,r^2\,coshx+sinhx) $,

$\Psi_{t3}(x)=(8\,a\,c\,A\,r^2\,cosh2x)+8\,coshx\,(-4\,a\,B\,+c\,a\,B\,\,r^2+2\,A\,c\,sinhx) $,

$\Psi_{t4}(x)=a\,B\,sinhx\,(1+c\,r^2\,sinh^2x)\,[\Psi_{t3}(x)+2\,B\,(5\,c-8\,a^2\,r^2+3\,c\,cosh2x)\,sinhx]$,

$\Psi_{t5}(x)=2\,A\,c+B\,(c+16\,a^2\,r^2)\,coshx-2\,c\,A\,cosh2x$,

$\Psi_{t6}(x)=\Psi_{t5}(x)-c\,B\,cosh3x+16\,a\,B\,sinhx-4\,a\,c\,r^2(2\,B\,sinhx+A\,sinh2x)$,

$\Psi_{d1}(x)=-4\,a^2\,r^2\,cosh2x\,(1+c\,r^2\,sinh^2x)-5\,a\,sinh2x+4\,a^2\,c\,r^4\,sinh^2{2x}$.

%%%%%%%%%%%%%%%%%%%%%%%%%%%%%%%%%%%%%%%%%%%%%%%%%%%%%%%%%%%%%%%%%%%%%%%%%%%%%%%%%%%%%%%%%%%%%%%%%%%%
\begin{figure}[h!] \centering
	\includegraphics[width=5cm]{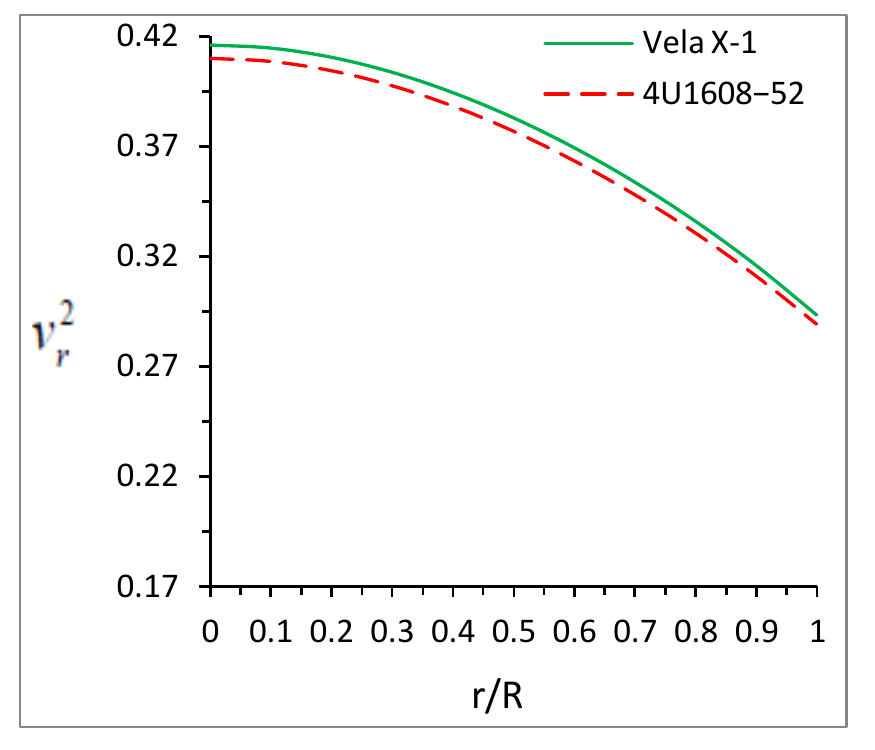}  \includegraphics[width=5cm]{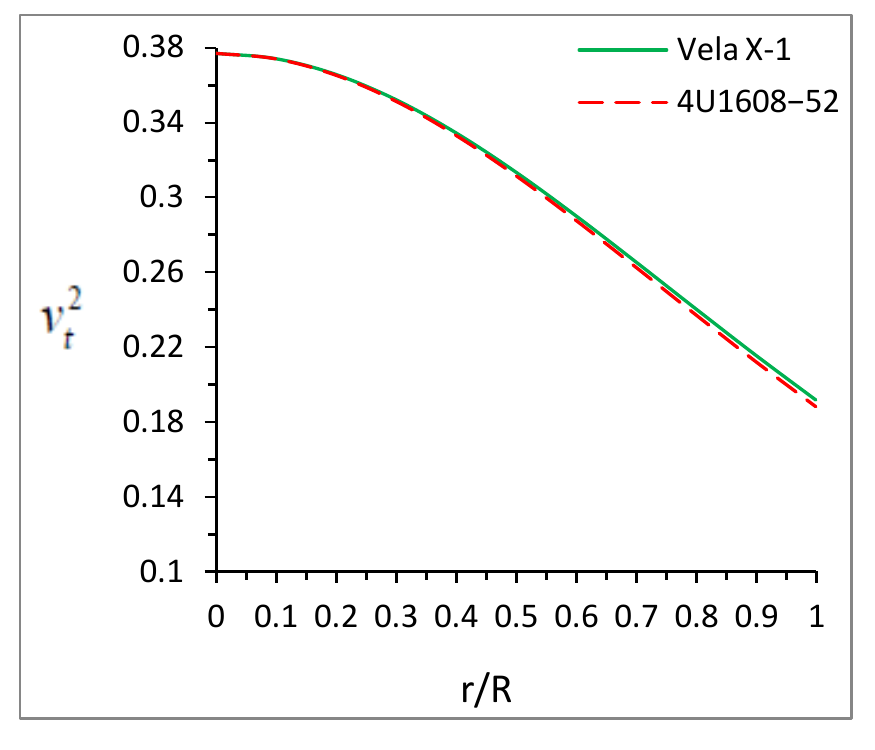}
	\caption{The behavior of $v^2_r$ (left panel) and $v^2_t$ (right panel) are shown versus $r/R$ for the anisotropic star Vela X-1 and 4U1608-52. The numerical values for plotting this graph are as follows: (i). $a=0.00033$, $b = 0.19$, $c = 0.2663$, $R= 9.56 km$ and $M=1.77 M_{\odot}$ for Vela X-1 (ii). $a = 0.000325$, $b = 0.1845$, $c = 0.276$, $R=9.528$ and $M=1.74 M_{\odot}$ for 4U1608-52. }.
	\label{12}
\end{figure}
%%%%%%%%%%%%%%%%%%%%%%%%%%%%%%%%%%%%%%%%%%%%%%%%%%%%%%%%%%%%%%%%%%%%%%%%%%%%%%%%%

%%%%%%%%%%%%%%%%%%%%%%%%%%%%%%%%%%%%%%%%%%%%%%%%%%%%%%%%%%%%%%%%%%%%%%%%%%%%%%%%%%%%%%%%%%%%%%%%%%%%
\begin{figure}[h!] \centering
	\includegraphics[width=5cm]{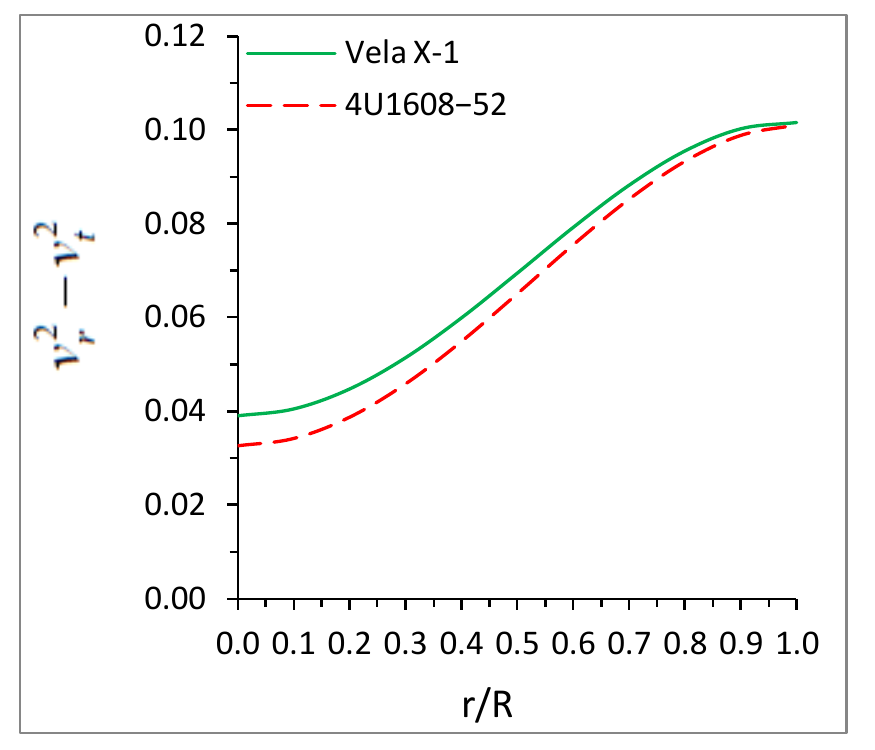}  \includegraphics[width=5cm]{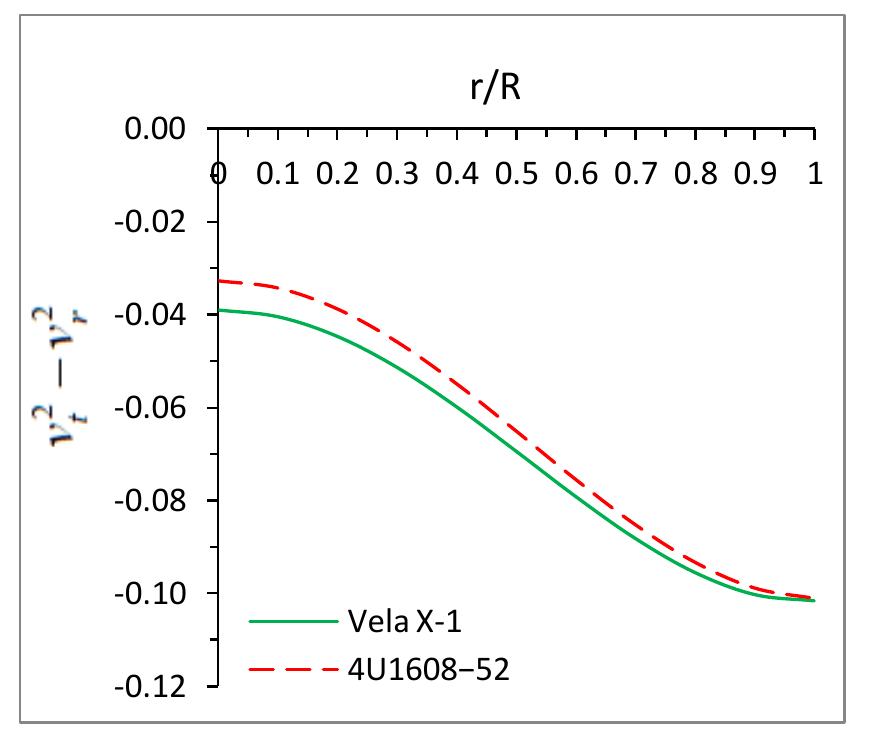}
	\caption{The behavior of $v^2_r-v^2_t$ (left panel) and $v^2_t-v^2_r$ (right panel) are shown versus $r/R$ for the anisotropic star Vela X-1 and 4U1608-52.For this graph we have employed numerical values for $a$, $b$, $c$, $A$ and $B$ same as used in Fig.(\ref{12})  (see table 1). }.
	\label{13}
\end{figure}
%%%%%%%%%%%%%%%%%%%%%%%%%%%%%%%%%%%%%%%%%%%%%%%%%%%%%%%%%%%%%%%%%%%%%%%%%%%%%%%%%
Fig. \ref{12} shows that both the radial and transverse velocities satisfy the causality conditions, i.e., both $v_r^2,\,v_t^2$ are less than unity and are monotonic decreasing functions of the radial coordinate.
We observe from Fig.(\ref{13}) that $v_{t}^2-v_{r}^2<0$ throughout the distribution thus indicating that our models are stable.

\subsection{Relativistic adiabatic index:}

\begin{figure}[h!] \centering
	\includegraphics[width=5cm]{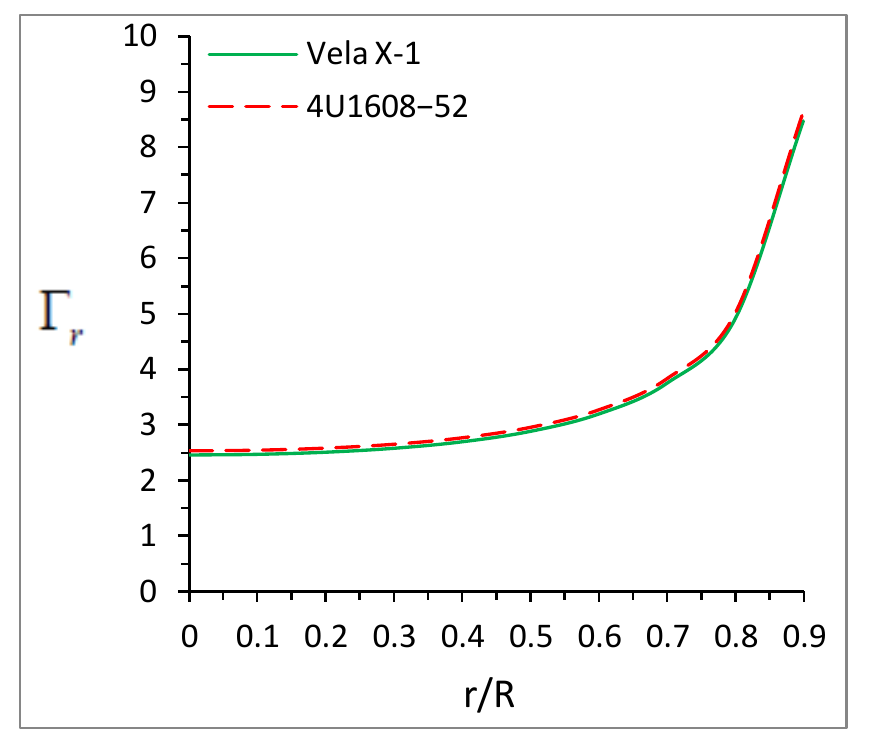}
	\caption{variation of $\Gamma_r$ versus $r/R$ is shown for the anisotropic stars Vela X-1 and 4U1608-52. For this graph we have employed numerical values for $a$, $b$, $c$, $A$ and $B$ same as used in Fig.(\ref{12}) and (\ref{13})  (see table 1).}.
	\label{14}
\end{figure}

A comprehensive discussion of the influence of pressure anisotropy and dissipation in collapsing, radiating fluids in the Newtonian and post-Newtonian limits is provided by Herrera and Santos\cite{hands}. Chandrasekhar showed that the ratio of the specific heats for an anisotropic fluid is given by
\begin{equation} \label{stable1}
\Gamma < \frac{4}{3} - \left[\frac{4}{3}\frac{p_r - p_t}{|{p_r}^\prime|r}\right]_{max}\end{equation}.
We note that the anisotropy increases the instability of the collapsing system when $p_r < p_t$. In the case of isotropic pressure, $p_r = p_t$ we obtain the classical Newtonian result, $\Gamma < \frac{4}{3}$ which is indicative of an unstable configuration. In the post-Newtonian approximation the unstable range of gamma is increased further due to relativistic corrections arising from the radial pressure increasing the effective density of the system as can be seen in the last term within the square brackets below
\begin{equation} \label{stable2}
\Gamma < \frac{4}{3} - \left[\frac{4}{3}\frac{p_r - p_t}{|{p_r}^\prime|r}+ \frac{1}{3}\kappa\frac{\rho p_r}{|{p_r}^\prime|}r\right]_{max}\end{equation}
It is possible that the anisotropy factor, $\Delta$ may change sign within the configuration. This would imply the existence of stable and unstable regions within the object which could lead to fragmentation of the sphere. From Fig. \ref{14} we observe that $\Gamma > \frac{4}{3}$ at each interior point of our models thus indicating that these models are stable.

%%%%%%%%%%%%%%%%%%%%%%%%%%%%%%%%%%%%%%%%%%%%%%%%%%%%%%%%%%%%%%%%%%%%%%%%%%%%%%%%%%%%%%%%%%%%%%%%%%%%
%%%%%%%%%%%%%%%%%%%%%%%%%%%%%%%%%%%%%%%%%%%%%%%%%%%%%%%%%%%%%%%%%%%%%%%%%%%%%%%%%

\subsection{Stability of static matter by Harrison-Zeldovich-Novikov criterion:}

In a recent paper, Singh et al.\cite{npg} employed the Harrison-Zeldovich-Novikov criterion to further investigate the stability of their models describing relativistic compact stars. In this formalism the configuration is stable only if the mass of the star is increasing with central density i.e. $dM/d\rho_0 > 0$ and unstable if $dM/d\rho_0 \le 0$.
Let us define the mass function of our static solution in terms of central density as,

\begin{equation}
M=\frac{R^3}{2}\,\frac{8\,\pi\,\rho_0\,sinh^2(aR^2+b)}{[3\,sinh^2b+8\,\pi\,\rho_0\,sinh^2(aR^2+b)]},
\end{equation}

after taking derivative of above equation with respect to $\rho_0$ we get,

\begin{equation}
\frac{dM}{d\rho_0}=\frac{R^2}{2}\,\frac{24\,\pi\,sinh^2b\,sinh^2(aR^2+b)}{[3\,sinh^2b+8\,\pi\,\rho_0\,sinh^2(aR^2+b)]^2},
\end{equation}

%%%%%%%%%%%%%%%%%%%%%%%%%%%%%%%%%%%%%%%%%%%%%%%%%%%%%%%%%%%%%%%%%%%%%%%%%%%%%%%%%%%%%%%%%%%%%%%%%%%%
\begin{figure}[h!] \centering
	\includegraphics[width=5.5cm]{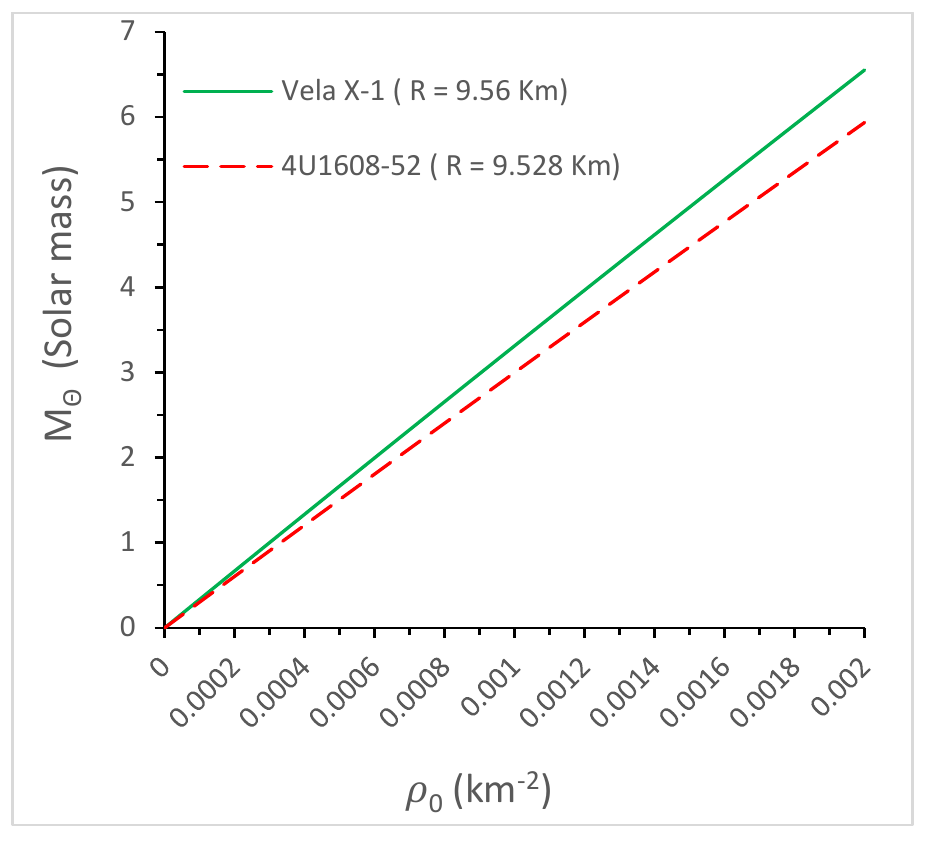}\includegraphics[width=5.5cm]{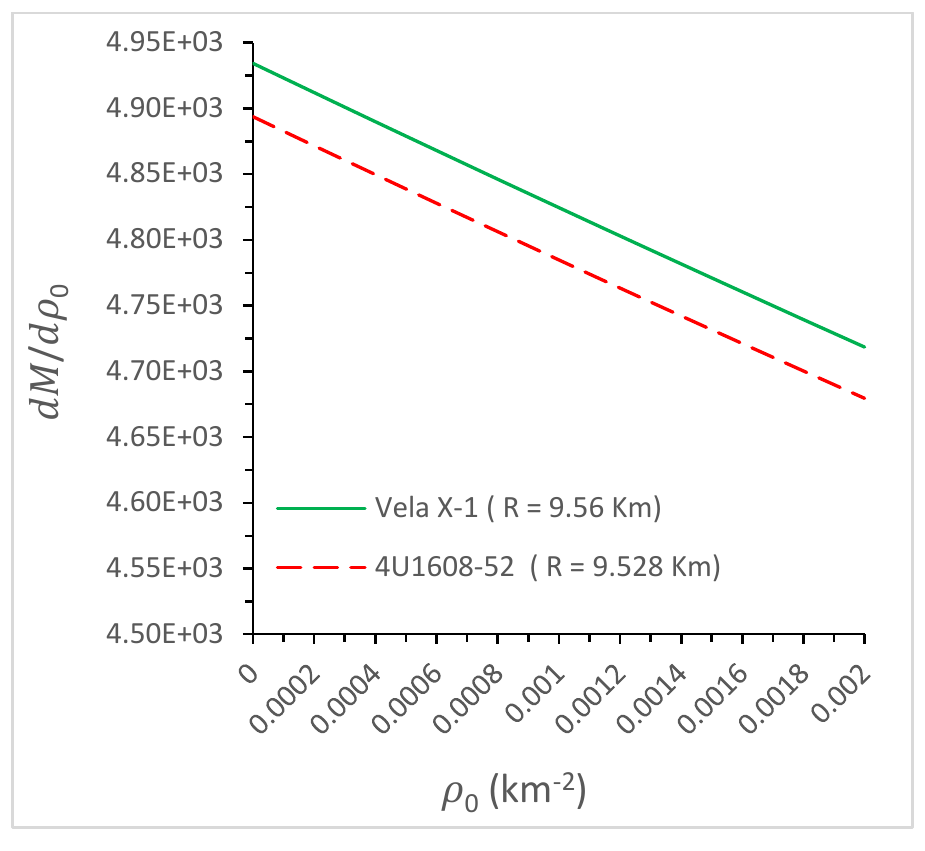}
	\caption{variation of Mass ($M_\odot$)(left panel) and $\frac{dM}{d\rho_0}$(right panel) versus central density $\rho_0 (0-2.6993\times10^{15} gm/cm^3) $  for the anisotropic star Vela X-1 and 4U1608-52. For this graph we have employed numerical values for $a$, $b$, $c$, $A$ and $B$ same as used in Fig.(\ref{13}) and (\ref{14})  (see table 1).}
	\label{15}
\end{figure}

%%%%%%%%%%%%%%%%%%%%%%%%%%%%%%%%%%%%%%%%%%%%%%%%%%%%%%%%%%%%%%%%%%%%%%%%%%%%%%%%%

We have plotted
$dM/d\rho_0 $ as a function of $\rho_0$ in Fig. \ref{15} (right panel) and it is clear that $dM/d\rho_0 > 0$ thus rendering our models stable.

\subsection{Energy conditions:}
The stellar configuration must satisfy the null energy condition (NEC), weak energy condition (WEC) and strong energy condition (SEC). These
conditions respectively are

\begin{equation}
NEC: \rho(r)\geq 0 ,
\end{equation}
\begin{equation}
WEC: \rho(r)-p_r(r) \geq  0~~ and
~~\rho(r)-p_t(r) \geq  0,
\end{equation}
\begin{equation}
SEC
: ~~\rho-p_r(r)-2p_t(r) \geq  0.
\end{equation}

%%%%%%%%%%%%%%%%%%%%%%%%%%%%%%%%%%%%%%%%%%%%%%%%%%%%%%%%%%%%%%%%%%%%%%%%%%%%%%%%%%%%%%%%%%%%%%%%%%%%
\begin{figure}[h!] \centering
	\includegraphics[width=5.5cm]{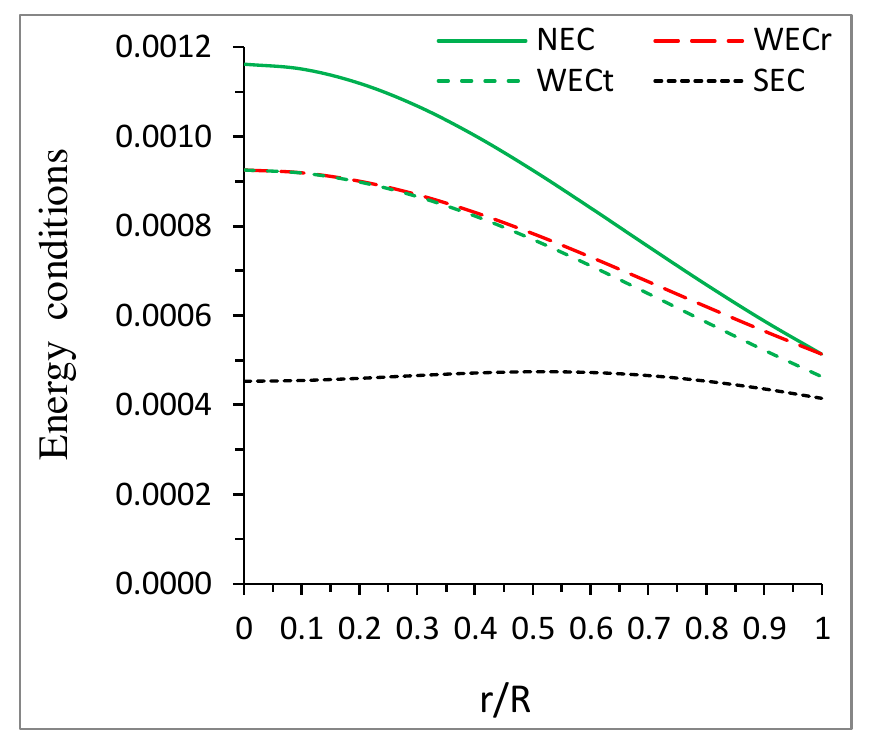}  \includegraphics[width=5.5cm]{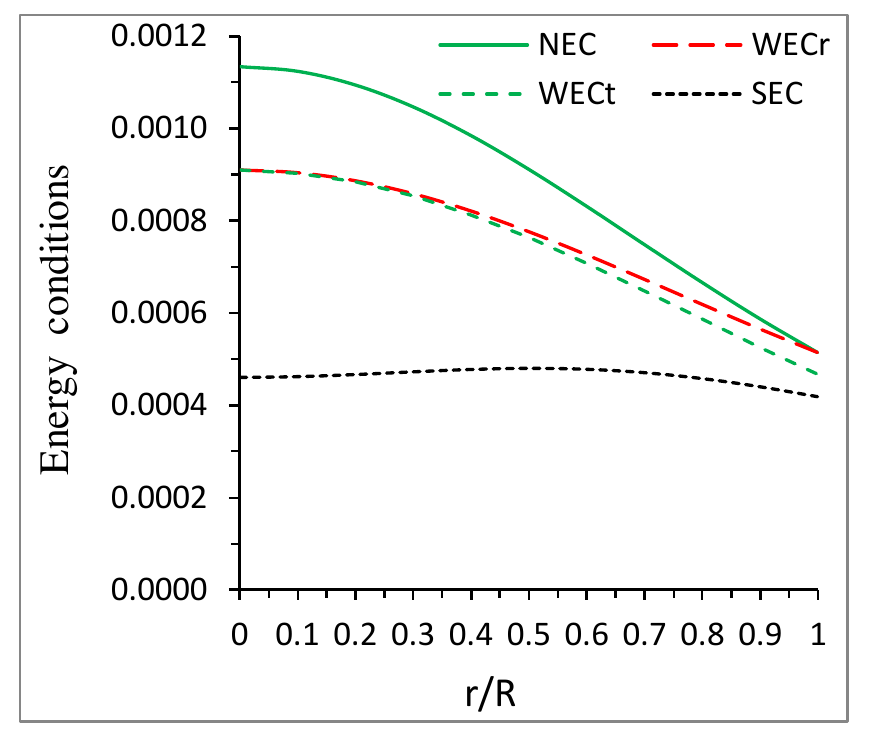}
	\caption{The NEC, WEC and SEC are plotted against $r/R$ for the anisotropic star Vela X-1 (left panel) and 4U1608-52 (right panel).The numerical values for plotting this graph are as follows:(i). $a=0.00033$, $b = 0.19$, $c = 0.2663$, $R= 9.56 km$ and $M=1.77 M_{\odot}$ for Vela X-1 (ii). $a = 0.000325$, $b = 0.1845$, $c = 0.276$, $R=9.528$ and $M=1.74 M_{\odot}$ for 4U1608-52.}.
	\label{ec}
\end{figure}
%%%%%%%%%%%%%%%%%%%%%%%%%%%%%%%%%%%%%%%%%%%%%%%%%%%%%%%%%%%%%%%%%%%%%%%%%%%%%%%%%
Fig. \ref{ec} clearly show that all the energy conditions are satisfied for each of our stellar models. We should point out that the cracking scenario derived by Abreu et al. \cite{Ab} in terms of the relative sound speeds within the fluid configuration requires that the strong and dominant energy conditions be satisfied.
\begin{table}
	\centering
	\caption{Numerical values of the parameters or constants for different compact stars}
	\label{Table1}
	\begin{tabular}{@{}lrrrrrr@{}} \hline
		$a(km^{-2})$ & $b(km^{-2})$ & $c$     & $A$     &$B(km^{-1})$  & Compact star \\ \hline
		0.00033     &  0.19        &  0.2663 &- 30.2557 & 30.1953     & Vela X-1  \\ \hline
		0.000325    & 0.1845       &0.276    &-31.1496 & 31.1140     & 4U1608-52 \\ \hline
		
	\end{tabular}
\end{table}

%%%%%%%%%%%%%%%%%%%%%%%%%%%%%%%%%%%%%%%%%%%%%%%%%%%%%%%%%%%%%%%%%%%%%%%%%%%%%%%%%
\begin{table}
	\centering
	\caption{Comparison between estimated mass and observed value of mass and radius for different compact stars\cite{GA}}
	\label{Table2}
	\begin{tabular}{@{}lrrrrrrr@{}} \hline
		$M/M_{\odot}$ & $R\,(Km)$ & $M/R$ & $M/M_{\odot}$&$R\,(Km)$ & Compact star \\
         $(estimated)$&$(estimated)$&$(estimated)$&$(observed)$&$(observed)$ & \\\hline
		 1.77 & 9.56   & 0.27276 & 1.77 $\pm$ 0.08 & 9.56 $\pm$ 0.08 & Vela X-1  \\ \hline
	     1.74 & 9.528  & 0.2690 & 1.74 $\pm$ 0.14 & 9.528 $\pm$ 0.15 & 4U1608-52 \\ \hline
		
	\end{tabular}
\end{table}

\begin{table}
	\centering
	\caption{Energy densities, central pressure and Buchdahl condition for different compact star candidates for the above parameter values of Tables 1 }
	\label{Table3}
	\begin{tabular}{@{}lrrrr@{}} \hline
		Central Density & Surface density & Central pressure & Buchdahl & Compact star  \\
		$gm/cm^{3} $ & $gm/cm^{3}$ & $dyne/cm^{2} $ & condition\\\hline
		1.5674$\times 10^{15} $ & 6.9285$\times 10^{14} $ & 2.8699$\times 10^{35} $ &$2M/R=0.5455 < 8/9$ & Vela X-1  \\ \hline
		1.5308$\times 10^{15} $ & 6.9467$\times 10^{14} $ & 2.7288$\times 10^{35}$ & $2M/R=0.5381 < 8/9$ & 4U1608-52  \\ \hline
		
	\end{tabular}
\end{table}

\subsection{Surface redshift:}
The effective mass of stellar configuration enclosed within radius $R$ is given by

\begin{equation}
M_{eff}=\frac{\kappa}{2}\int_0^{R}\rho\, r^{2}dr=\frac{R}{2}\,\left[\frac{cR^2\,sinh^2 (aR^2+b)}{1+cR^2\,sinh^2 (aR^2+b)}\right]  \label{eq30},
\end{equation}

The surface redshift in terms of compactness $u=M/R$ is defined as,

\begin{equation}
z_s=\frac{1-[1-2u]^{\frac{1}{2}}}{[1-2u]^{\frac{1}{2}}}=\sqrt{1+cR^2\,sinh^2(aR^2+b)}-1  \label{eq30}
\end{equation}

%%%%%%%%%%%%%%%%%%%%%%%%%%%%%%%%%%%%%%%%%%%%%%%%%%%%%%%%%%%%%%%%%%%%%%%%%%%%%%%%%%%%%%%%%%%%%%%%%%%%
\begin{figure}[h!] \centering
	\includegraphics[width=5.5cm]{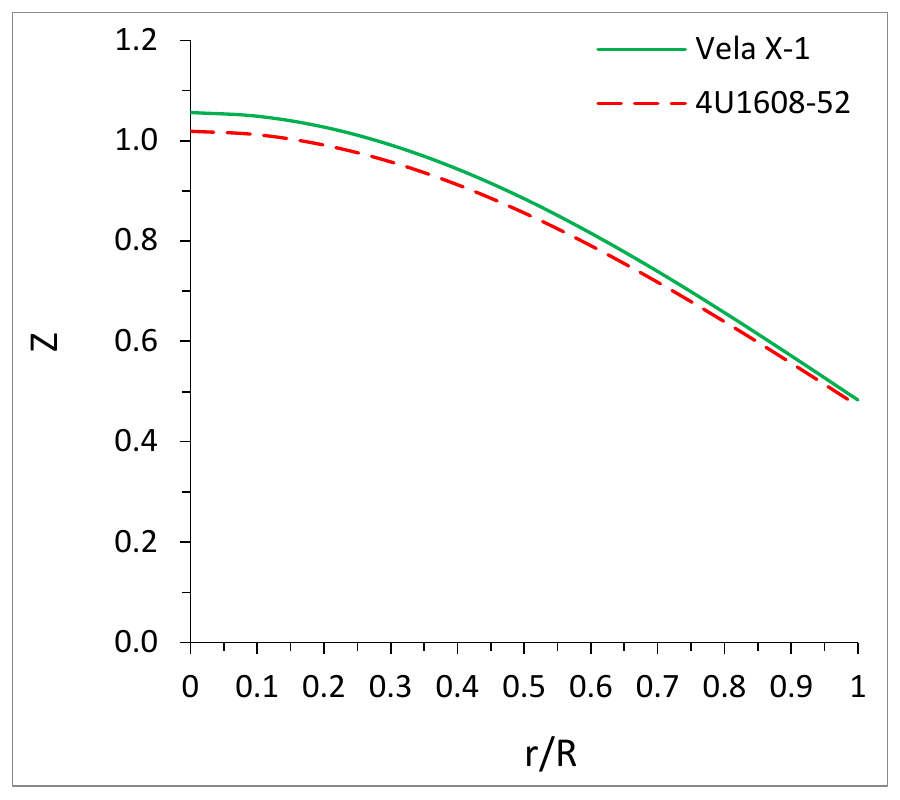}
	\caption{The variation of redshift ($z$) against $r/R$ for the anisotropic star Vela X-1  and 4U1608-52.The numerical values for plotting this graph are as follows: (i). $a=0.00033$, $b = 0.19$, $c = 0.2663$, $R= 9.56 km$ and $M=1.77 M_{\odot}$ for Vela X-1 (ii). $a = 0.000325$, $b = 0.1845$, $c = 0.276$, $R=9.528$ and $M=1.74 M_{\odot}$ for 4U1608-52.}.
	\label{red1}
\end{figure}
%%%%%%%%%%%%%%%%%%%%%%%%%%%%%%%%%%%%%%%%%%%%%%%%%%%%%%%%%%%%%%%%%%%%%%%%%%%%%%%%%
\section{Discussion}
\noindent In this work we presented a non singular solution of Einstein's field equations for anisotropic fluid distribution satisfying Karmarkar's
condition. From fig. 1 it can be seen that $e^{\lambda}$ and $e^{\nu}$ are well behaved throughout the star hence there is no signature
problem. The radial pressure $p_{r}$ and tangential pressure $p_{t}$ are positive throughout the distribution and decreasing
radially outwards as can be seen from fig. 2. Fig. 3 shows the variation of density throughout the distribution which is
positive and decreases monotonically from the center towards the surface of the star. It can be observed from fig. 4 that the anisotropy parameter $\Delta$ is zero
at the centre and increases radially outwards. It is clear from fig. 5 that the anisotropic force, hydrostatic force and gravitational
force all are well behaved throughout the interior of the stellar configuration. Fig. 6, it can be seen that the square of sound velocity is positive and less than 1. Hence the model satisfy
all the physical plausibility conditions.

\noindent The stability of the stellar configuration has been checked via Herrera's cracking method, the trend in the adiabatic index,  $\Gamma$ and the Harrison-Zeldovich-Novikov stability criterion.
From fig. 10 clearly shows that the stellar configuration satisfies null energy condition,
weak energy condition and strong energy condition. Fig. 11 shows surface redshift is also well behaved. The compactification factor
$M/R$ is displayed in table 2. From the table it is clear that our model is in
good agreement with the most recent observational data of pulsars given by Gangopadhyay {\em et. al.}\cite{GA}.

\noindent Hence our model is stable, satisfies all the physical plausibility conditions and is useful to describe compact stars
like Vela X-1 and 4U1608-52.

\noindent In summary, we began with a static spherically symmetric metric and imposed the Karmarkar condition which is a necessary condition for a metric to be of embedding class 1. The Karmarkar condition reduces the problem of finding solutions of the Einstein field equations to a single-generating function. If we further require that the interior matter distribution of the star be described by a perfect fluid then the only possible configuration is the Schwarzschild uniform density sphere. It is well-known that this solution leads to infinite propagation speeds within the stellar core. In order to generate a physically viable model of a compact star, we chose a metric potential which is singularity-free and well-behaved throughout the stellar interior. We are in a position to integrate the Karmarkar condition to produce a closed form for the remaining metric potential. We have demonstrated that this solution satisfies all the requirements for a stable, matter configuration in hydrostatic equilibrium. Our model displays a great degree of robustness in approximating observable compact objects.

\section*{Acknowledgments}
S. K. Maurya acknowledge continuous support and encouragement from the administration of University
of Nizwa. BSR is thankful to IUCAA Pune, for facilities provided to him where the part of work was carried out.

\end{document}